\documentclass[prd,twocolumn,superscriptaddress,showpacs,preprintnumbers,amsmath,amssymb,nofootinbib]{revtex4}

\usepackage{dcolumn}
\usepackage{bm}
\usepackage[latin1]{inputenc}
\usepackage[spanish,english]{babel}
\usepackage{amsfonts}
\usepackage{amssymb}
\usepackage{graphicx}

\newcommand{\be}{\begin{equation}}
\newcommand{\ee}{\end{equation}}
\newcommand{\bea}{\begin{eqnarray}}
\newcommand{\eea}{\end{eqnarray}}

\begin{document}
\title{{\bf Insensitivity of Hawking radiation to an invariant Planck-scale cutoff}}
\author{Iván Agulló}\email{ivan.agullo@uv.es}
 \affiliation{ {\footnotesize Physics Department, University of
Wisconsin-Milwaukee, P.O.Box 413, Milwaukee, WI 53201 USA}}\affiliation{ {\footnotesize Departamento de Física Teórica and
IFIC, Centro Mixto Universidad de Valencia-CSIC.
    Facultad de Física, Universidad de Valencia,
        Burjassot-46100, Valencia, Spain. }}
\author{José Navarro-Salas}\email{jnavarro@ific.uv.es}
\affiliation{ {\footnotesize Departamento de Física Teórica and
IFIC, Centro Mixto Universidad de Valencia-CSIC.
    Facultad de Física, Universidad de Valencia,
        Burjassot-46100, Valencia, Spain. }}

\author{Gonzalo J. Olmo}\email{olmo@iem.cfmac.csic.es }
\affiliation{\footnotesize Instituto de Estructura de la Materia,
CSIC, Serrano 121, 28006 Madrid, Spain}
\author{Leonard Parker}\email{leonard@uwm.edu}
\affiliation{ {\footnotesize Physics Department, University of
Wisconsin-Milwaukee, P.O.Box 413, Milwaukee, WI 53201 USA}}

\date{June 29, 2009}

\begin{abstract}

A disturbing aspect of Hawking's derivation of black hole radiance
is the need to invoke extreme conditions for the quantum field
that originates the emitted quanta. It is widely argued that the
derivation requires the validity of the conventional relativistic
field theory to arbitrarily high, trans-Planckian scales. We stress in
this note that this is not necessarily the case if the question is
presented in a covariant way. We point out that  Hawking
radiation is immediately robust against an invariant Planck-scale
cutoff. This important feature
of Hawking radiation is relevant for a quantum gravity theory  that
preserves, in some way, the Lorentz symmetry.
\end{abstract}

\pacs{04.62+v,04.70.Dy}

\maketitle

The Hawking effect \cite{hawking} plays a pivotal role in the interplay between quantum mechanics and general relativity and, hence, it is of special relevance in any proposal for a quantum gravity theory. The original derivation of Hawking is based on the general  framework
of particle creation on curved spacetimes, first developed in a cosmological setting in
\cite{parker} (see also \cite{parker-toms, birrel-davies}).
The derivation considers the  propagation of modes 
that represent particles in the asymptotically flat regions;
the first at early times before a dust cloud has begun to
collapse, and the second at late times long after it
has collapsed to form a black hole 
as seen by a distant observer.
In short, the
expansion of a field in two different sets of  modes,
$u^{in}_j(x)$ (that are positive frequency on past null infinity)
and $u^{out}_j(x)$ (that are positive frequency on future 
null infinity)
leads to a relation for the corresponding
creation and annihilation operators:  $ a_i^{out}=\sum_j (\alpha^*
_{ij}a_j^{in}-\beta ^*_{ij}a_j^{in \dagger})$.
When the
coefficients $\beta _{ij}$ do not vanish,  the ``in'' and ``out'' vacuum 
states do not coincide and, therefore, the
number of particles measured in the $i^{th}$ mode by an ``out''
observer  in the ``in'' vacuum state, is given by
 $\langle  N_i \rangle = \sum_k|\beta_{ik}|^2$.
 For a Schwarzschild black hole, one obtains \cite{hawking} for the average number of particles observed at late times 
in the state in which no particles are present at early times
(we omit angular quantum numbers) 
\be \label{integralbeta}\langle
N_w \rangle = \int_0^{+\infty} dw' |\beta_{ww'}|^2 \ee 
where the
beta coefficients, up to a transmission amplitude factor and a 
trivial phase
are given by \be \label{betaww'}
\beta_{w,w'}=\frac{1}{2\pi\kappa}\sqrt{\frac{w'}{w}}\frac
{\Gamma(1+\kappa^{-1}wi)}{ (-\kappa^{-1}w')^{1+\kappa^{-1}wi}}
 \ , \ee
where $\kappa$ is the surface gravity. Since these coefficients 
behave like $1/\sqrt{w'}$ for large $w'$,
the integral (\ref{integralbeta}) diverges. This is naturally
interpreted as the fact that the total number of created quanta is
infinite, as corresponds to a finite steady rate of  emission. The
steady rate can be easily obtained from (\ref{integralbeta})  and
turns out to be thermal 
\be \label{dotN} \langle \dot{N}_w
\rangle =\frac{1}{2\pi}\frac{1}{e^{2\pi \kappa^{-1}w}-1} \ . \ee
However, there is a  disturbing point in this derivation. One
needs to perform an unbounded integration in the 
frequencies $w'$ to obtain the steady thermal rate of radiation 
\cite{Parker77, Wald84, jacobson9193, fabbri}.
Any out-going Hawking
quanta at infinity will have an  exponentially
increasing frequency as they are propagated backwards in time 
to reach the near-horizon region.

A cutoff in the frequencies $w'$ of order of 
the Planck length (we take units with $c=1$) would require
that we consider only early-time frequencies satisfying
\be\label{cutoffw}w' < \ell_P^{-1} \ , \ee where $\ell_P$ is the Planck length.
This will change completely the
Hawking effect. It will introduce a damping time-dependent factor in
formula (\ref{dotN}). The Hawking radiation is then converted into
a transient phenomena (see, for instance, \cite{agullo-navarro-olmo-parker} and also \cite{gil}).

However, as first shown in \cite{fredenhagen-haag},  it is possible  to re-derive the Hawking radiation from a different perspective. In this derivation it is just the universal Hadamard short distance behavior of the two-point function
for all physically allowed  states near horizon, namely
\be G(x_1, x_2) \approx  \frac{\hbar}{4\pi^2 \sigma} \ , \ee
 where $\sigma$ is the squared geodesic distance between $x_1$ and $x_2$, that is responsible for the steady thermal emission. 
 A somewhat related
  approach was developed  in \cite{agullo-navarro-olmo-parker, agullo-navarro-olmo}. The mean  number operator at late times can be expressed, in general, as \cite{agullo-navarro-olmo-parker, agullo-navarro-olmo}
  \begin{equation}\label{eq:N-eps}
\langle N_{i}\rangle = \hbar^{-1} \int_\Sigma
d\Sigma_1 ^\mu d\Sigma_2 ^\nu
[u^{out}_{i}(x_1){\buildrel\leftrightarrow\over{\partial}}_\mu
][u^{out*}_{i}(x_2){\buildrel\leftrightarrow\over{\partial}}_\nu
] G(x_1, x_2).
\end{equation}
After some algebra, one arrives at the expression \bea
\langle N_w\rangle = -\frac{1}{4\pi^2
w}\int_{-\infty}^{0}dU_1 dU_2
\frac{e^{-iw(u(U_1)-u(U_2))}}{(U_1-U_2-i\epsilon)^2}  \ , \eea
where $U$ is the null Kruskal coordinate $U=-\kappa^{-1}e^{-\kappa
u}$ and $u=t-r^*$ is the corresponding retarded time of a
Schwarzschild black hole. The double integral above is divergent,
but this divergence is expected due to the infinite number of
quanta emitted in the  infinite amount of time involved in the
formula. Restricting the computation to the mean particle number
per unit time one gets the finite thermal result \bea
\label{N-epsI-}\langle \dot N_w\rangle &=& -\frac{1}{4\pi^2
w}\frac{d}{du}\int_{-\infty}^{0}dU_1 dU_2
\frac{e^{-iw(u(U_1)-u(U_2))}}{(U_1-U_2-i\epsilon)^2}\nonumber \\
&=& \frac{1}{2\pi}\frac{1}{e^{2\pi \kappa^{-1}w} -1}\ . \eea
Again, the disturbing point in the above derivation is that a
cutoff in distances requiring that 
\be \label{cutoffU}(U_1 - U_2)^2 >
\ell^2_P \ , \ee turns the otherwise steady Hawking radiation into a transient phenomenon. 
One notices immediately that the common
point in the cutoff (\ref{cutoffU}) and that of (\ref{cutoffw}) is
that both are not Lorentz-invariant. 
Since we have put an upper limit, $w' \sim 1/\ell_P$,
on the early-time frequencies, the ``in'' modes remaining after
this amputation are not sufficient to generate the radiated ``out'' modes at late times.
This produces the described decay of Hawking radiation with time as a consequence of breaking the principle
of relativity by means of a non-invariant cutoff. \\

It is possible, however,  to introduce a cutoff in an invariant way. On dimensional grounds, one can demand that the two-point function
$G(x_1, x_2)$ that appears in our integrals does not exceed
the inverse of Newton's constant  
\be
|G(x_1, x_2)| <  \hbar \ell_P^{-2} \equiv G_N^{-1}\ . 
\label{conditionG}
\ee It is not difficult to show, as
we will see, that this condition translates into a restriction in
the integration range of the $U_1, U_2$ coordinates
 in (\ref{N-epsI-}) given by
\be \label{conditionU} (U_1-U_2)^2 > \ell_P^2\kappa^2U_1U_2/4\pi^2 \ . \ee
The factor $\kappa^2U_1U_2$ on the right hand side of (\ref{conditionU}) is absent in Eq. (\ref{cutoffU}). This factor is required to have an invariant cutoff for all locally inertial observers and immediately ensures the robustness of Hawking radiation.

An understanding of how  (\ref{conditionU}) follows from  (\ref{conditionG}) can be obtained in a simple way by considering the Unruh effect \cite{unruh}.
A detector held at constant $r$ just outside the horizon behaves like a uniformly accelerated detector in Minkowski space (equivalence principle). 
The thermal radiation detected by the accelerated observer can be related
to the Hawking emission.
The detector will have some internal energy states $|E\rangle$ and it can interact with the field by absorbing or emitting  quanta.
The interaction can be modeled in the standard way by coupling the field $\phi(x)$ along the detector trajectory $x=x(\tau)$ ($\tau$ is the detector proper time) to some operator $m(\tau)$ acting on the internal detector eigenstates
\be g \int d\tau\, m(\tau)\Phi(x(\tau)) \ , \ee
where $g$ is the strength of the coupling.
The probability for  the detector to make the
transition from $|E_i\rangle$ to $|E_f\rangle$ is given by the expression
$P(E_i \to E_f) = g^2 |\langle E_f |m(0)|E_i\rangle|^2 F(\Delta E)$,
 where  $F(\Delta E)$ is the so-called response function
\be \label{FE}F(\Delta E)=
\int_{-\infty}^{+\infty}d\tau_1d\tau_2
e^{-i\Delta E \Delta \tau/ \hbar}\langle 0_M|
\Phi(x(\tau_1))\Phi(x(\tau_2)) |0_M\rangle \ ,  \ee where $\Delta \tau= \tau_1 -\tau_2$.
For a massless field the Wightman two-point function in (\ref{FE}), where $|0_M\rangle$ is the Minkowski vacuum, is given by
\be \langle 0_M|
\Phi(x_1)\Phi(x_2)|0_M\rangle= -\frac{\hbar}{4\pi^2[(\Delta t -i\epsilon)^2 -(\Delta \vec{x})^2]}  \ . \ee
For trajectories having a proper-time translational symmetry under
$\tau \to \tau + \tau_0$, it is natural to consider the constant
transition probability per unit proper time
and the corresponding response rate per unit proper time
\be \label{RRF}
{\dot F(\Delta E)} = \int_{-\infty}^{+\infty}d\Delta
\tau e^{-i\Delta E \Delta \tau/\hbar}\langle 0_M|
\Phi(x(\tau_1))\Phi(x(\tau_2))|0_M\rangle  \ . \ee
Both the inertial detector and the uniformly accelerated detector possess proper-time translational symmetry.
 For an inertial detector trajectory, the response rate is given by
\bea \label{rateFinertial}{\dot
F(\Delta E)}&=& -\int_{-\infty}^{+\infty}d\Delta \tau e^{-i\Delta E\Delta
\tau/\hbar}\left[
\frac{\hbar}{4\pi^2(\Delta \tau - i\epsilon)^2}\right]\nonumber \\ &=& -\frac{\Delta E}{2\pi} \theta(-\Delta E)\ , \eea
in agreement with the principle of relativity. If the detector's initial state is the  ground state $E_i =E_0$, then $\Delta E >0$ and the probability for an inertial detector  to be excited is exactly zero, irrespective of the velocity of the detector.  (When $\Delta E <0$ the result is non-vanishing and this leads to the expected non-zero probability for the spontaneous decay $E_i \to E_f<E_i$.)

For a uniformly accelerated trajectory in Minkowski spacetime
\begin{equation} \label{accelerated}
t =\frac{1}{a} \sinh{a\tau} \ , \ x= \frac{1}{a} \cosh{a\tau} \ ,
\end{equation} where $a$ is the acceleration, the response function is then
\be \label{rateF00}{ F(\Delta E)}=
\int_{-\infty}^{+\infty}d\tau_1 d\tau_2 e^{-i\frac{\Delta E\Delta\tau}{ \hbar}}
\frac{-\hbar({a}/{2})^2}{4\pi^2\sinh^2
\left[ \frac{a}{2}( \Delta \tau
-i\epsilon)\right]} \ . \ee
The corresponding response rate function
turns out to be ${\dot F(\Delta E)}= (\Delta E /2\pi)(e^{2\pi \Delta E/\hbar a}
-1)^{-1}$,
which implies, via the
detailed balance relation, 
${\dot P}(\Delta E) = {\dot P}(-\Delta E)e^{-2\pi \Delta E/a \hbar}$,  
that a uniformly accelerated
observer in Minkowski space feels himself immersed in a thermal
bath at the temperature $k_B T = \frac{a \hbar}{2\pi}$. 

Performing the change of variable 
\be \label{Utau} U\equiv t -x =
-a^{-1}e^{-a\tau} \ , \ee 
one can rewrite the integral (\ref{rateF00}) in the form 
\be \label{rateF1b} F(\Delta E)=-
\int_{-\infty}^{0}dU_1dU_2 e^{-i\Delta E\Delta \tau/\hbar}
\frac{\hbar}{4\pi^2 (U_1 - U_2 -i\epsilon)^2} \ . \ee 
The time derivative of this
expression is exactly the same (up to the factor 
$1/\hbar w$)
as (\ref{N-epsI-}) obtained before in computing the expectation value of the number
operator in the Hawking effect 
(identifying the acceleration $a$ with the surface 
gravity $\kappa$ and the coordinate
$U$ with the corresponding Kruskal coordinate).
It is now easy to see that the invariant cutoff condition \be
\label{2pcutoff} \left|\frac{\hbar}{4\pi^2[(\Delta t)^2 -(\Delta
\vec{x})^2]}\right|< G_N^{-1} \  \ee on the accelerated trajectory
(\ref{accelerated}) becomes \be \label{sinh}
\frac{\hbar(\frac{a}{2})^2}{4\pi^2\sinh^2\frac{a}{2}\Delta \tau }
< G_N^{-1} \ . \ee 
Expanding the denominator of (22) to lowest order in $\Delta \tau$ and using (19) to express $(\Delta \tau)^2$ in terms of $(\Delta U)^2\equiv (U_1-U_2)^2$, it is straightforward to show that this inequality is equivalent to (\ref{conditionU}). This confirms our statement that (\ref{conditionG}) implies (\ref{conditionU}).\\

The natural question now is to see if the invariant cutoff
suffices to preserve the bulk of the Hawking effect. The answer is
in the affirmative, but to see this requires an additional step \cite{agullo-navarro-olmo-parker, agullo-navarro-olmo}. Let us use again the Unruh effect to illustrate the argument.  We want to take advantage of the fact that there is a state of the field, $|0_A\rangle$,  for which the response function of the accelerated detector vanishes for $\Delta E >0$ 
\bea
{ F_A(\Delta E >0 )}& =&
\int_{-\infty}^{+\infty}d\tau_1 d\tau_2 e^{-i\Delta E \Delta \tau} \times \\ & & \langle 0_A|
\Phi(x(\tau_1))\Phi(x(\tau_2))| 0_A\rangle =0\nonumber . \eea

Taking this into account, it is possible to obtain an equivalent expression for the response function of the uniformly accelerating detector in the Minkowski vacuum, $|0_M\rangle$, by subtracting the previous quantity from the right-hand-side of equation (\ref{FE})
 \bea
\label{calibration}  &&{ F(\Delta E >0 )} =
\int_{-\infty}^{+\infty}d\tau_1 d\tau_2 e^{-i\Delta E \Delta \tau}
\times  \\ &&[\langle 0_M|
\Phi(x(\tau_1))\Phi(x(\tau_2))|0_M\rangle - \langle 0_A|
\Phi(x(\tau_1))\Phi(x(\tau_2))| 0_A\rangle] \nonumber  .\eea
This expression presents several advantages over (\ref{FE}).
It explicitly shows that the difference between two-point correlation functions of the field in the vacuum states $|0_M\rangle$ and $|0_A\rangle$ is at the root of a non-vanishing response function. (Notice that although the integral of 
$\langle 0_A |\Phi(x(\tau_1))\Phi(x(\tau_2))| 0_A\rangle$ in the response function is zero, the correlation function itself is not zero.)
Moreover, the integrand is now a smooth and symmetric function,
thanks to the universal short-distance behavior of the two-point
functions. Thus, the usual ``$i\epsilon-$prescription'' in the two-point functions
is now redundant and can be omitted. Additionally, expression (\ref{calibration}) shows a remarkable fact when an invariant cut-off is considered. It manifestly produces a vanishing result in the limit $a\to0$, respecting in that way the principle of relativity that we want to preserve.

Now, one can
consistently implement the invariant and universal cutoff
condition
\be \label{condition1}|\langle 0_M|
\Phi(x(\tau_1))\Phi(x(\tau_2))|0_M\rangle| < G_N^{-1} \ , \ee and
\be |\langle 0_A| \Phi(x(\tau_1))\Phi(x(\tau_2))| 0_A\rangle| <
G_N^{-1} \ \ee
in (\ref{calibration}). The first inequality is
equivalent to (\ref{sinh}) and the second one  to $\Delta \tau^2 >
\ell_P^2/4\pi^2 $. Moreover, both inequalities are essentially
equivalent  since all quantum states (in particular $|0_M\rangle$
and $|0_A\rangle$) have the same short distance behavior, as is seen explicitly from the short distance asymptotic form of (\ref{sinh}).

In the black hole case, the same argument can be applied for the computation of
the mean particle number \cite{agullo-navarro-olmo-parker, agullo-navarro-olmo}, and $G(x_1,x_2)$ in equation (\ref{eq:N-eps}) can be substituted by

\be  G(x_1,x_2)-\langle out|\Phi(x_1) \Phi(x_2)|out\rangle \ ,\ee
where $|out\rangle$ is, as usual, the vacuum state defined by the modes $u^{out}_j(x)$. This leads to an expression for the mean particle number per unit time
 \bea
\label{Nbh}\langle \dot N_w\rangle &=& -\frac{1}{4\pi^2
w}\frac{d}{du}\left[\int_{-\infty}^{0}dU_1 dU_2
\frac{e^{-iw(u(U_1)-u(U_2))}}{(U_1-U_2)^2}\right. \nonumber \\&-& \left.\int_{-\infty}^{+\infty}du_1 du_2
\frac{e^{-iw(u_1-u_2)}}{(u_1-u_2)^2}\right] \ , \eea
where now we want to restrict the ranges of integration, so
$(U_1-U_2)^2 > \ell_P^2\kappa^2U_1U_2/4\pi^2$ and $(u_1 - u_2)^2 > \ell_P^2/4\pi^2$. The explicit evaluation of these integrals, with the corresponding bounds for $(U_1-U_2)^2$ and $(u_1 - u_2)^2$, leads to
\be
\langle \dot N_w\rangle \approx  \frac{1}{2\pi}\frac{1}{e^{2\pi \kappa^{-1}w}-1} - \frac{\kappa\ell_P}{96\pi^4(w/\kappa)}+ O(\kappa \ell_P)^3 \ . \ee
For black hole radii much bigger than the planck length ($\kappa\ll \ell_P^{-1}$) and for reasonable values of the frequency,  the correction terms are negligible, which shows the irrelevance of ultra-high energy physics in the derivation of the Hawking effect.

In summary, we have shown that a universal invariant cutoff condition for two-point functions is able to preserve the bulk of the thermal Hawking radiation.


\noindent { \bf Acknowledgements} This
work has been supported by grant FIS2008-06078-C03-02. L.P. and I.A. have been partly
supported by NSF grants PHY-0071044 and PHY-0503366 and by
a UWM RGI grant. G.O. thanks MICINN for a JdC contract.

\end{document}